\documentclass[12pt,a4paper]{iopart}
\usepackage{iopams}
\usepackage{graphicx}

\newcommand{\transpose}{^{\mathsf{T}}}

\begin{document}

\title[Modelling of Oscillations in 2d Echo-Spectra of the Fenna-Matthews-Olson Complex]{Modelling of Oscillations in Two-Dimensional Echo-Spectra of the Fenna-Matthews-Olson Complex}
\author{Birgit Hein$^{1}$, Christoph Kreisbeck$^{1}$, Tobias Kramer$^{1,2}$ and Mirta Rodr\'iguez$^{3}$}
\address{
$^1$Institut f\"ur Theoretische Physik, Universit\"at Regensburg, 93040 Regensburg, Germany\\
$^2$Department of Physics, Harvard University, Cambridge, Massachusetts 02138, USA\\
$^3$Instituto de Estructura de la Materia CSIC, C/ Serrano 121, 28006 Madrid, Spain
}
\date{\today}
\ead{birgit.hein@physik.uni-regensburg.de,tobias.kramer@mytum.de}
%\linenumbers

\begin{abstract}
Recent experimental observations of time-dependent beatings in the two-dimensional echo-spectra of light-harvesting complexes at ambient temperatures have opened up the question whether coherence and wave-like behaviour plays a significant role in photosynthesis. We perform a numerical study of the absorption and echo-spectra of the Fenna-Matthews-Olson (FMO) complex in {\it chlorobium tepidum} and analyse the requirements in the theoretical model needed to reproduce beatings in the calculated spectra. The energy transfer in the FMO pigment-protein complex is theoretically described by an exciton Hamiltonian coupled to a phonon bath which account for the pigments electronic and vibrational excitations respectively. We use the hierarchical equations of motions method to treat the strong couplings in a non-perturbative way. We show that the oscillations in the two-dimensional echo-spectra persist in the presence of thermal noise and static disorder.
\end{abstract}

\maketitle

\section{Introduction}
\label{sec:intro}

Light harvesting complexes are the part of photosynthetic systems that channel energy from the antenna to the reaction centre. One of the most studied complexes is the Fenna-Matthews-Olson (FMO) complex \cite{Ishizaki2010a,Fenna1975,Olson62} which is part of the photosynthetic apparatus of green sulphur bacteria. The observation of long lasting beatings, up to $1.8$~ps at $T=77$~K in time-resolved two-dimensional (2d) spectra \cite{Collini2010a,Engel2007a,Panitchayangkoon2010} has generated enormous interest in theoretical modelling the process of energy transfer through light-harvesting complexes. Recently molecular-dynamics simulations started to develop a more microscopic picture of the dynamics \cite{Olbrich2011,Shim2011}, but an atomistic and fully quantum-mechanical model of the whole process reproducing the experimentally obtained spectra is not available. Therefore theoretical models treat the light-harvesting pigment-protein complex as an exciton system coupled to a bath which accounts for the vibrational modes of the pigments. 

The choreography of the exciton-energy transfer can be elucidated with 2d echo-spectroscopy using three laser pulses which hit the  sample within several femtoseconds time spacings. The experimentally measured 2d echo-spectra of the complex show a variety of beating patterns with oscillation periods roughly consistent with the differences in eigenenergies of the excited system obtained from fits to the absorption and 2d echo-spectra \cite{Engel2007a}. At long delay times between pump and probe pulses the system displays relaxation to the energetically lowest lying exciton eigenstates as expected from the interactions between the electronic degrees of freedom and the vibrational modes of the complex which drive the system into a state of thermal equilibrium \cite{Brixner2005}.

One main goal of the research in this field is to clarify the role of a disordered and fluctuating environment in the energy-transfer efficiency. Theoretical quantum-transport studies show that the energy transfer in these systems results from a balance between coherent and dissipative dynamics \cite{Mohseni2008,Caruso2009}. Calculations at room temperature also show that the optimal energy-transport efficiencies appear as a compromise between coherent dynamics and thermal fluctuations \cite{Kreisbeck2011}.
  
Another open question in the theoretical study of these systems is the description of experimental observables, i.e.\ the absorption and the 2d echo-spectra. Most of the theoretical studies analyse the beatings in the population dynamics and compare them with the ones observed in the experimentally measured 2d echo-spectra. However, the 2d echo-spectrum reflects different information than the population dynamics and needs to be calculated and analysed separately in detail. 
In principle techniques such as quantum state and process tomography made out of a sequence of 2d echo-spectra can be used to map out the complete density matrix \cite{Yuen-Zhou2011}. In contrast to energy-transfer efficiency studies where an initial excitation enters the complex at specific sites close to the antenna, in 2d echo-spectra the whole complex is simultaneously excited. Also the two-exciton manifold yields prominent contributions to the signal, resulting in negative regions in the 2d echo-spectra.
﻿
The non-pertubative calculation of 2d echo-spectra presents a considerable computational challenge due to the presence of two excitons giving raise to excited state absorption and the requirement to consider ensemble averages over differently orientated complexes with slightly varying energy levels. Previous calculations have employed Markovian approximations \cite{Cho2005,Bruggemann2007} or exclude the double exciton manifold \cite{Sharp2010}. 
In addition the systematic study of beatings in a series of 2d echo-spectra requires effective means to calculate a huge number of such spectra.
So far no theoretical method has been able to describe the long-lasting beatings in the time-resolved 2d spectra \cite{Chen2011,Cho2005,Sharp2010}. One possible explanation for the persistence of long coherence times has been the sluggish absorption of the reorganisation energy by the molecule, which requires theoretical descriptions that go beyond the Markovian approximation and the rotating wave approximation \cite{Thorwart2009}. The hierarchical equations of motions (HEOM), first developed by Tanimura and Kubo \cite{Tanimura1989} and subsequently refined in \cite{Yan2004, Xu2005, Ishizaki2005, Tanimura2006}, show oscillations in the dynamics of the exciton populations that persist even at temperature $T=300$~K \cite{Ishizaki2009d,Ishizaki2009e}. The HEOM include the reorganisation process in a transparent way and are directly applicable for computations at physiological temperatures. A recent calculation at temperature $77$~K of the 2d echo-spectra with the HEOM method has been performed by Chen \etal \cite{Chen2011}, which does not display the strong beatings observed experimentally. Chen \etal conclude that the agreement between theory and experiments needs to be improved. Here, we analyse under what conditions the theoretical calculations give better agreement with the experimental results. We calculate the absorption and 2d echo-spectra in a temperature range between $77$~K to $277$~K using the HEOM and investigate the role of pigment and protein vibrations in the calculated spectra. We systematically discuss the parameters needed to obtain results that agree better with the experimentally measured spectra. We discuss the role of static disorder for the shapes of the peaks and the appearance of beatings in the time-resolved 2d echo-spectrum and their robustness against temperature changes and static disorder.

The paper is organised as follows: the exciton model and the HEOM are introduced in sec.~\ref{sec:model}. The absorption spectra and the line shapes obtained from the HEOM model of the FMO complex are analysed in sec.~\ref{sec:1d} where we compare different methods for performing rotational averages. In sec.~\ref{sec:2d} we discuss the 2d spectrum and analyse how disorder affects it. Furthermore we analyse the persistence of beatings in the spectra as a function of the delay time for several temperatures and with and without static disorder.

\section{Exciton model of the FMO complex}
\label{sec:model}

The FMO complex described by Fenna, Matthews and Olson \cite{Fenna1975, Olson62} is a pigment-protein complex that channels the energy in the photosynthetic apparatus in green sulphur bacteria. The FMO complex is arranged as a trimer with the different subunits interacting weakly with each other. We thus restrict our study to a single subunit which contains eight bacteriochlorophyll molecules (BChls) \cite{Tronrud2009,Ben-Shem2004}. The BChls pigments, which are wrapped in a protein environment, can be electronically excited and are coupled due to dipole-dipole interactions. The BChls form a network that guides the energy from the antenna to the reaction centre. Since the eighth BChl is only loosely bound it usually detaches from the others when the system is isolated from its environment to perform experiments  \cite{Tronrud2009, SchmidtamBusch2011}. Therefore, we do not take the eighth BChl into account and describe the energy transfer using the seven site Frenkel exciton model \cite{Leegwater1996, Ritz2001, May2004}. The single exciton manifold is given by
\begin{equation}\label{eq:FrenkelHam}
  H_{\rm e}^{\rm 1ex}=\sum_{k=1}^{N_{\rm sites}}\epsilon_k|k\rangle\langle k| 
+\sum_{k>l}J_{kl} \left( |k\rangle\langle l|+|l\rangle\langle k| \right)
\end{equation}
with $N_{\rm sites}=7$. The state $|k\rangle$ corresponds to an electronic excitation of BChl$_k$ with site energy $\epsilon_k$. Dipole-dipole couplings between different chromophores yield the inter-site couplings $J_{kl}$, which lead to a delocalisation of the exciton over all BChls. The site energies of the BChls $\epsilon_k=\epsilon_k^0+\lambda_k$ consist of `zero-phonon energies' $\epsilon_k^0$, and a term due to the reorganisation Hamiltonian $H_{\rm reorg}=\sum_{k=1}^{N_{\rm sites}}  \lambda_k \left|k\right\rangle \left\langle k\right|$. The reorganisation energy $\lambda_k=(\sum_i\hbar\,\omega_i\,q_i^2/2)_k $ denotes the energy which is removed from the system to adjust the vibrational coordinates to the new equilibrium value \cite{May2004,Ishizaki2009d,Yang2002}. Here, $q_i$ denotes a dimensionless displacement of the $i$th oscillator measured in units of $\sqrt{\hbar/(m_{i}\omega_{i})}$, where $m_{i}$ and $\omega_{i}$ are mass and frequency of the $i$th oscillator. The vibrational modes of the BChl are modelled as a set of harmonic oscillators $H_{\rm phon}=\sum_{i}\hbar\omega_i b_i^\dag b_i$. The vibrational environment couples linearly to the exciton system $H_{\rm ex-phon}=\sum_{k=1}^{N_{\rm sites}} \left( \sum_i \hbar \omega_i q_i(b_i+b_i^\dag)\right)_k \left|k\right\rangle\left\langle k\right|$ and induces fluctuations in the site energies. We neglect correlations between the vibrations at different sites and model the coupling strength by a structureless spectral density
\begin{equation}
\label{eq:spectral-density}
J(\omega)=2\lambda\frac{\omega \gamma}{\omega^2+\gamma^2},
\end{equation}
where we assume identical couplings $\lambda_k=\lambda$ at each site. 
This choice of spectral density facilitates the use of the HEOM method since the time-dependent bath correlations then assume an exponential form. 
For the FMO complex a variety of spectral densities are discussed in the literature, where some are a parametrisation of experimental fluorescence spectra \cite{Adolphs2006a}, while others are extracted from molecular dynamics simulations \cite{Olbrich2011a}.
We use the model Hamiltonian for {\it chlorobium tepidum} given in Ref.~\cite{Cho2005}. 
The Hamiltonian originates from previous calculations in \cite{Vulto1998,Vulto1999}. Cho et al.\ \cite{Cho2005} changed the coupling constant between BChl 5 and BChl 6 and obtained new site energies from a fit to experimental results. The Hamiltonian of Ref.~\cite{Cho2005} has been used in previous studies of 2d echo-spectra \cite{Cho2005,Sharp2010,Chen2011} and facilitates comparisons between the different approaches.
In the site basis $\left\{\left| k\right\rangle\right\}$ the matrix elements read
\begin{equation}
\label{eq:H}
H_{\rme}^{\rm 1 ex}=\left(\begin{array}{rrrrrrr}
 280 &  -106   &  8  &  -5  &   6  &  -8  &  -4\\
-106 &   420  &  28  &   6  &   2  &  13  &   1\\
   8 &    28  &   0 & -62   & -1  &  -9  &  17\\
  -5 &     6  & -62  & 175  & -70  & -19 &  -57\\
   6 &     2  &  -1 &  -70  & 320  &  40 &   -2\\
  -8  &   13  &  -9  & -19  &  40 &  360 &  32\\
  -4  &    1 &   17 &  -57 &   -2 &   32 &  260\\
\end{array}
\right)+\lambda_S \mathbf{1}
\end{equation}
in units of cm$^{-1}$, where we have added an energy shift $\lambda_S=12075$~cm$^{-1}$ to the diagonal elements of the exciton Hamiltonian $H_{e}^{\rm 1 ex}$ to shift the positions of the eigenenergies to the frequency range of the figures in Ref.~\cite{Cho2005}.  We set the parameters of the spectral density to $\lambda=35$~cm$^{-1}$ and $\gamma^{-1}=50$~fs. 
These parameters are chosen to provide best agreement to the smooth part of the spectral density in Ref.~\cite{Adolphs2006a}, where the low frequency component of the spectral density was fitted to fluorescence line narrowing spectra \cite{Wendling2000}. Since a single Lorentz-Drude peak is not sufficient to accurately reproduce the smooth part of the spectral density of Ref.~\cite{Adolphs2006a}, we choose the parameters to obtain agreement within a factor of $1.2$ to $1.5$ in the energy range from $100$ to $150$~cm$^{-1}$, which corresponds to the typical difference between the exciton energies.

The HEOM approach rewrites time non-local effects into a hierarchy of coupled equations of motion for a set of auxiliary matrices $\sigma^{\vec{n}}$. The superscript $\vec{n}$ is a vector of $N_{\rm sites}$ entries, where $n_{k}$ gives the order to which the bath of BChl$_k$ is taken into account. The reduced density matrix $\rho(t)=\tr_{\rm phonon}R(t)$ describing the exciton system is obtained after tracing out the vibrational degrees of freedom. The time evolution of the reduced density matrix is given by
\begin{equation}
\label{eq:drho}
\fl
\frac{d}{dt}\rho\left(t\right)=-\rmi{\cal L}_{\rme}\rho\left(t\right) +\sum_{k=1}^{N_{\rm sites}}\Phi_k\sigma^{\vec{e}_k}\left(t\right)
,\end{equation}
with 
\begin{equation}
\label{eq:dsigma}
\fl
 \frac{d}{dt}\sigma^{\vec{n}}\left(t\right)
=-\left(\rmi{\cal L}_{\rme}+\sum_{k=1}^{N_{\rm sites}}\gamma n_k\right)\sigma^{\vec{n}}\left(t\right)
+ \sum_{k=1}^{N_{\rm sites}} \Phi_k\sigma^{\vec{n}+\vec{e}_k}\left(t\right) + \sum_{k=1}^{N_{\rm sites}} n_k\Theta_k\sigma^{\vec{n}-\vec{e}_k}\left(t\right).
\end{equation}
${\cal L}_{\rm e}$ is the Liouville operator of the coherent exciton dynamics and we define $\vec{e}_{k}$ as the $k$th unit vector with $N_{\rm sites}=7$ entries. The operators $\Phi_k$  and $\Theta_k$ are defined by their action on a test operator $A$
\begin{eqnarray}
\Phi_k A &=\rmi V_k^\times A, \\
\Theta_k A &=\rmi \left(\frac{2\lambda_k}{\beta\hbar^{2}} V_k^\times A 
-\rmi\frac{\lambda_k\gamma}{\hbar}\left(\left|k\right\rangle\left\langle k\right| A+A \left|k\right\rangle\left\langle k\right|\right)\right)
,\end{eqnarray}
% %
where $V_k^\times A=[|k\rangle\langle k|,A]$. The auxiliary matrices are initialised to $\sigma^{\vec{n}}\left(0\right)=0$ for $\vec{n}\neq\vec{0}$ and $\sigma^{\vec{0}}\left(t\right)=\rho\left(t\right)$ \cite{Ishizaki2009d}. 
The hierarchy is truncated for sufficiently large $N_{\rm max}=\sum_{k=1}^{N_{\rm sites}} n_k$, where the diagonal coupling in (\ref{eq:dsigma}) becomes dominant. For the parameters used in the calculation, reorganisation energy $\lambda=35$~cm$^{-1}$ and phonon relaxation time scale $\gamma^{-1}=50$~fs, we verified that the relative differences in the interesting energy region of the absorption spectrum using $N_{\rm max}=4$ versus $N_{\rm max}=12$ are below $0.13$~percent. In the 2d echo-spectrum we compared the rephasing stimulated emission diagram for $N_{\rm max}=4$ and $N_{\rm max}=12$ and found a difference of less than $0.07$~percent of the amplitude. We therefore consider a truncation at $N_{\rm max}=4$ sufficient to reach converged results. The derivation of the hierarchy stated in (\ref{eq:drho}) and (\ref{eq:dsigma}) relies on a high temperature approximation. For a temperature of $77$~K the high temperature approximation is not valid and low temperature correction terms have to be included \cite{Ishizaki2009e}. Within the low temperature correction we replace 
\begin{eqnarray}\label{eq:lowtemp}
 {\cal L}_{\rm e}\rightarrow& {\cal L}_{\rm e}-\sum_{k=1}^{N_{sites}}\frac{2 \lambda_k}{\beta\hbar^2}\frac{2
 \gamma}{\nu_1^2-\gamma^2}V_k^\times V_k^\times, \nonumber \\
 {\Theta}_k\rightarrow& {\Theta}_k-\frac{2 \lambda_k}{\beta\hbar^2}\frac{2
 \gamma^2}{\nu_1^2-\gamma^2}V_k^\times,
\end{eqnarray}
where $\nu_1=2\pi /\beta\hbar$ is the first bosonic Matsubara frequency following \cite{Ishizaki2009e}. Agreement with \cite{Nalbach2011} and proper thermalization to the Boltzmann distribution \cite{Kreisbeck2011} confirms that the first order correction in (\ref{eq:lowtemp}) is indeed sufficient. A further improvement of the low temperature correction can be attained by including higher Matsubara terms. Recently, also a Pad\'e spectrum decomposition has been proposed \cite{Hu2010,Hu2011} which shows faster convergence than the Matsubara decomposition.

The numerical evaluation of the HEOM requires to propagate a large number of auxiliary matrices. Although the number of auxiliary matrices can be reduced by a scaled HEOM approach \cite{Shi2009a, Zhu2011} as well as by advanced filtering techniques \cite{Shi2009a}, the HEOM are computationally very demanding, which sets limits to treatable system sizes. For 2d echo-spectroscopy one has to propagate the set of auxiliary matrices not only for a huge number of time steps (typically of the order of $10^5$), but also the dimension of the auxiliary matrices gets enlarged since excited state absorption requires to extend the exciton Hamiltonian to the two exciton manifold. We overcome the computational limitation by using a GPU (Graphics Processing Unit) implementation where the computation time goes down with the number of cores on a single GPU \cite{Kreisbeck2011}. For example on a single NVIDIA C2050 graphics board we obtain a $450$ fold speedup compared to an equivalent single-core CPU implementation. With our efficient GPU algorithm the computation time for a 2d spectrum is reduced to 50 minutes on a single GPU. 

\section{Absorption spectra of the FMO complex}
\label{sec:1d}

Spectroscopy provides a tool that gives direct insight into the energy states of a quantum system. Absorption spectroscopy is used to identify the energy eigenstates and their interaction strengths with the radiation field, expressed in terms of dipole moments. The energy eigenstates and the dipole moments are commonly identified with the peak position and the peak areas respectively. The spectral peaks are broadened due to the interaction of the energy eigenstates with the remaining degrees of freedom of the system. For an ensemble of light-harvesting complexes the coupling of the electronic dynamics to the vibrational modes and the presence of fluctuations in the energy levels, due to e.g. the motion of the surrounding proteins, broadens the absorption peaks. 

To calculate absorption-spectra we have to expand the system and add the ground state $\left|0\right\rangle$ to our basis. The Hamiltonian is thus extended to an $8 \times 8$ block matrix
\begin{equation}\label{eq:1d-H}
 H_{\rme}^{1d}=
\left(\begin{array}{cc}
 H_{\rme}^{\rm gs} & 0 \\  
 0 & H_{\rme}^{\rm 1 ex}
\end{array}\right)
,
\end{equation}
where we have chosen the zero exciton energy $H_{\rme}^{\rm gs}=0$ and $H_{e}^{\rm 1 ex}$ is the $7 \times 7$ matrix defined in (\ref{eq:H}).

Using the same block structure as in (\ref{eq:1d-H}) the dipole matrix that governs the photon-exciton conversion is constructed as 
\begin{equation}
\mu=
\left(\begin{array}{cc}
 0 &  \mu_{\rm 1ex}^{-} \\  
 \mu_{\rm 1ex}^{+} & 0
\end{array}\right)
\label{eq:1d-mu}
,\end{equation}
where $\mu_{\rm 1ex}^{+}={\mu_{\rm 1ex}^{-}}\transpose$ and the zero to single excitation dipole block
\begin{equation}
\mu^{+}_{\rm 1ex}=\sum_{k=1}^{N_{\rm sites}} \mu_k\left|k\right\rangle\left\langle 0\right|
\label{eq:mu_1ex}
\end{equation}
is defined using the seven dipole moments $\mu_k$ that give the excitation of BChl$_k$ by the laser pulse.

The absorption spectrum is calculated as
\begin{equation}
I_{\rm 1d}(\omega)= \left\langle \int_0^\infty dt\, \rme^{\rmi \omega t} \tr\left( \mu\left(t\right) \mu (0) \rho_{0}\right) \right\rangle_{\rm rot}
\label{eq:1d-I}
,\end{equation}
where $\rho_{0}=\left|0\right\rangle \left\langle 0\right|$ is the density operator for no excitation and we assume $\delta$-shaped laser pulses. The different methods for performing the rotational average denoted by $\langle\cdot\rangle_{\rm rot}$ are detailed below.
The time evolution of the dipole operators in (\ref{eq:1d-I}) is given by $\mu\left(t\right)=\rme^{\rmi H_{e}^{1d}t/\hbar}\mu\rme^{-\rmi H_{e}^{1d}t/\hbar}$.

In the FMO complex the seven BChl molecules have a fixed orientation with respect to each other. Within this configuration each BChl has a dipole vector $\vec{d}_k$ that characterises its interaction with light. We consider the dipole vectors of the BChls to be along the connection line of two nitrogen atoms N$_{B}-$N$_{D}$ \cite{Adolphs2008} whose positions we obtain from the protein data bank \cite{Tronrud2009,TronrudPDB}. The norm of $\vec{d}_k$ accounts for the strength of the dipole and we assume $\left|\vec{d}_k\right|=d$ since all BChl molecules are identical. For the simulations we choose $d=1$. Given a polarisation direction of the laser $\vec{l}=(\sin\beta\cos\gamma,\sin\beta\sin\gamma, \cos\beta)$, where $\gamma \in [0,2\pi)$ and $\beta \in [0,\pi]$, the dipole moments in (\ref{eq:mu_1ex}) read
\begin{equation}
\mu_k=\vec{d}_k\cdot\vec{l}
.\end{equation}
Typical experiments are performed using spectroscopy of a sample solution and many FMO complexes are illuminated at the same time. Each of them is randomly orientated with respect to the electric field in the laser pulse. If we assume a coherent dynamics expression (\ref{eq:1d-I}) is readily evaluated in eigenbasis as
\begin{eqnarray}
\fl
\left\langle \tr\left( \mu\left(t\right) \mu (0) \rho_{0}\right)\right\rangle_{\rm rot} 
&=&\frac{1}{4\pi}\sum_{j=1}^{7}\rme^{-\rmi E_j t/\hbar}\int_{0}^{2\pi} d\gamma \int_{0}^{\pi}d\beta\, \sin \beta\, \left(\sum_{k=1}^{7}\vec{d}_{k}\cdot\vec{l}\,v_{k}^{j}\right)^{2}\\
&=&\sum_{j=1}^7 \bar{\mu}^2_{E_j} \rme^{-\rmi E_j t/\hbar}
,\end{eqnarray}
where $E_{j}$ is the $j$th eigenenergy and $v_{k}^{j}=\langle E_j| k\rangle$ denotes the overlap of the $k$th state in site basis with the $j$th eigenstate. The integrationally averaged dipole moments in eigenbasis are
\begin{eqnarray}
\bar{\mu}_{E_j}^2
&=\frac{1}{4\pi}\int_{0}^{2\pi}d\gamma \int_{0}^{\pi}d\beta\, \sin\beta\,\left(\sum_{k=1}^{7}\vec{d}_{k}\cdot\vec{l}\,v_{k}^{j}\right)^{2}	\nonumber \\
&=\frac{1}{3}\sum_{k,m=1}^{7}v_{k}^{j}v_{m}^{j}\vec{d}_{k}\cdot\vec{d}_{m}\label{eq:preavdip}
.\end{eqnarray}
Thus, we obtain the integrationally averaged squares of dipole moments in the eigenbasis $\bar{\mu}^{2}/d^{2}=\left\{0.25,\ 0.64,\  0.39,\  0.11,\  0.62,\ 0.12,\  0.20\right\}$, where the eigenenergies increase from left to right. For an appropriate $d$, these values are in rough agreement with the averaged dipole moments obtained from a fit to the experimental spectrum  $\mu_{\rm FIT}^{2}=\left\{49,\ 87,\ 73,\ 31,\ 82,\ 24,\ 36\right\}$~debye$^2$ \cite{Cho2005}.

However, we can also read the result as $\bar{\mu}_{E_j}^{2}=\frac{1}{3} \sum_{k,m=1}^{7}v_{k}^{j}v_{m}^{j}\sum_{i=1}^{3}\vec{d}_k\cdot\vec{l}_{i}\,\vec{d}_m\cdot\vec{l}_{i}$, where the three polarisation vectors $\vec{l}_{i}$ are orthogonal to each other. This now implies that the rotational average can be performed using three orthogonal directions for $\vec{l}$ instead of sampling over the whole unit sphere. Note that this calculation is valid only if the interaction of system and bath does not interfere with the rotational average. 

In the following we discuss how the line-shapes of the peaks depend on the theoretical method and approximations commonly used for performing the time-dependent propagation in (\ref{eq:1d-I}) before we return to the different methods of performing the rotational average.

\subsection{Absorption spectra obtained by different propagation methods}
\label{sec:results1d}

We calculate the absorption spectra of the FMO complex at the temperature $T=77$~K.
We propagate the exciton-bath system during $2048$~fs with a numerical time step of $2$~fs using the HEOM with reorganisation energy $\lambda=35$~cm$^{-1}$ and a time scale for the bath correlations given by $\gamma^{-1}=50$~fs. Each HEOM propagation takes only 1 second on a GPU. 

\begin{figure}[t]
\includegraphics[width=\textwidth]{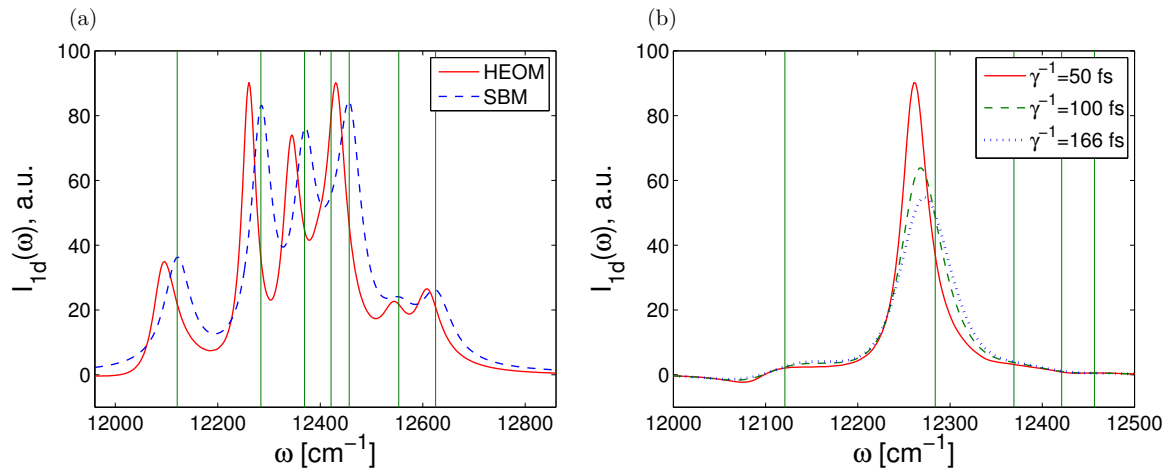}
\caption{(a) Absorption spectra $I_{\rm 1d}(\omega)$ (\ref{eq:1d-I}) for the FMO complex calculated at $T=77$~K
using the HEOM and the secular Born-Markov (SBM) approximation (without Lamb-shift) for  $\gamma^{-1}=50$~fs, $\lambda=35$~cm$^{-1}$ and $\mu_{\rm FIT}$ . The vertical lines correspond to the eigenvalues of the exciton Hamiltonian (\ref{eq:H}). (b) Absorption lines of the second eigenstate calculated using the HEOM for different values of $\gamma^{-1}$ in the spectral density (\ref{eq:spectral-density}) with fixed $\lambda=35$~cm$^{-1}$.}
\label{fig:1d-both}
\end{figure}

In figure~\ref{fig:1d-both}~(a), we compare the results obtained for the absorption spectrum calculated using the HEOM and the secular Born-Markov approximation (SBM) following Refs.~\cite{Rebentrost2009,Kreisbeck2011}. The line shapes for each eigenenergy can be calculated in the eigenbasis
$I_{j}=\int_0^{\infty} \rme^{\rmi\omega t}\left\langle\tr\left(\mu_{E_j}\left(t\right) \mu_{E_j}\left(0\right)\rho_{0}\right)\right\rangle_{\rm rot}$.
In the SBM simulation the complete spectrum is identical to the sum of the seven positive absorption peaks (not shown). Due to the rotating wave approximation the terms $\tr\left(\mu_{E_j}\left(t\right) \mu_{E_l}\left(0\right)\rho_{0}\right)$ for $j\neq l$ are zero. For the HEOM, these terms have non-vanishing contributions and the sum of the line shapes  $I_k$  do no longer yield the spectrum. Thus we have to calculate the line shape of the $j$th exciton peak  as $I_j=\int_0^{\infty}\rme^{\rmi\omega t}\left\langle\tr\left(\sum_{i=1}^{N_{\rm sites}} \mu_{E_i}\left(t\right) \mu_{E_j}\left(0\right)\rho_{0}\right)\right\rangle_{\rm rot}$. 
Including these terms leads to more complicated line shapes with shoulders and negative contributions as shown in figure~\ref{fig:1d-both}~(b) for the peak of the second eigenstate. We find that the more complicated line shapes arise from the non-secular approximations. Comparisons of different shapes of the spectrum obtained by different propagation methods and different spectral densities have been presented in \cite{Schroder2006,Chen2009} for the B850 ring of the light-harvesting complex LH2.

We observe in figure  \ref{fig:1d-both} (a) that the energies for the HEOM peaks are shifted compared to the peaks obtained in the SBM approximation. Since we do not take into account the time-dependent Lamb  shift in the SBM approximation (see eq.~(8) in Ref.~\cite{Rebentrost2009}) the corresponding SBM peaks are located at the exciton energies of Hamiltonian (\ref{eq:H}), marked by the vertical lines. 
Note that the eigenenergies of Hamiltonian (\ref{eq:H}) include the reorganisation energy $\lambda$ added to the ``bare exciton-energies''. Due to the dissipation of the reorganisation energy within the vibrational environment, taken into account by the HEOM \cite{Ishizaki2009d}, the peaks of the absorption spectrum shift to lower energies. For a fast phonon relaxation $\gamma^{-1}\rightarrow 0$ (Markov limit) the reorganisation energy dissipates instantaneously and the peaks are shifted to lower energies by approximately $\lambda$ and then coincide with the bare exciton energies. Figure  \ref{fig:1d-both}(b) displays how the bath-correlation time affects the peak shapes and positions. For longer phonon relaxation times, the reorganisation energy dissipates slowly during the exciton dynamics and the peaks stay closer to the eigenenergies of (\ref{eq:H}).

\begin{figure}[t]
\centering
\includegraphics[width=\textwidth]{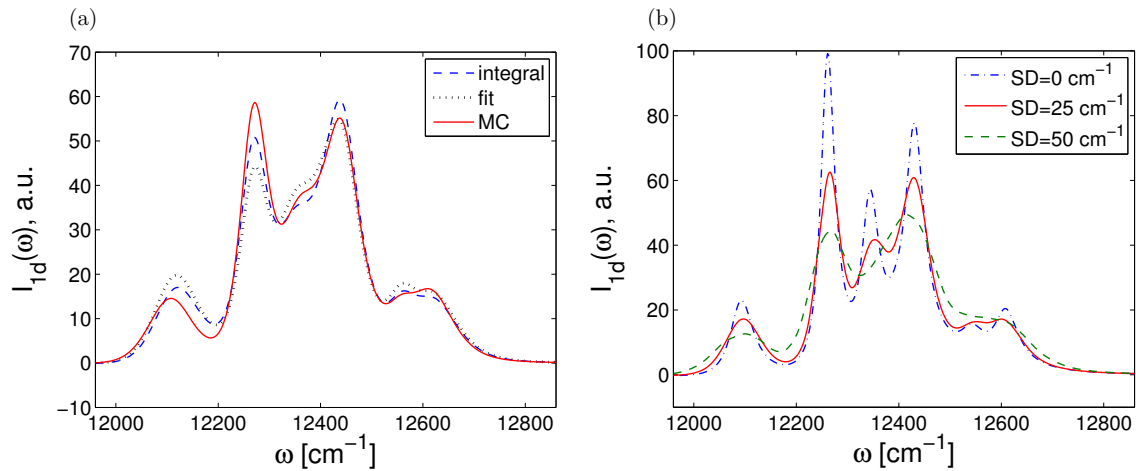}
\caption{(a) Absorption spectra $I_{\rm 1d}(\omega)$ of the FMO complex calculated using the HEOM with $N_{\rm max}=8$ at $T=77$~K, $\lambda=35$~cm$^{-1}$ and $\gamma^{-1}=166$~fs. We compare different methods of performing the rotational average in the calculations. The dotted line results from using the fitted values $\mu_{\rm FIT}$, the dashed line takes the values $\bar{\mu}$. The solid line denotes a Monte Carlo (MC) average over $300$ samples. The spectra are normalised with respect to the area under the graph. (b) Comparison of the absorption spectra for $\gamma^{-1}=50$~fs for different values of static Gaussian disorder, which is characterised by its standard deviation (SD). In all cases the rotational average was performed with MC using 300 realisations.}
\label{fig:1d-averages}
\end{figure}

While the spectra shown above are calculated using fitted dipole moments from \cite{Cho2005}, we will now analyse the changes due to rotational averaging on the absorption spectrum. In figure~\ref{fig:1d-averages}~(a) we show the results for the  spectrum using the HEOM for $\lambda=35$~cm$^{-1}$ and $\gamma^{-1}=166$~fs. We compare different methods of rotational average discussed in the previous section, that is, a single propagation from integrationally averaged dipole moments using $\bar{\mu}$ eq.~(\ref{eq:preavdip}), a Monte Carlo (MC) set of 300 orientations and $\mu_{\rm FIT}$. The results using three propagations with perpendicular laser directions $xyz$ are not shown since the curve is almost identical to the MC spectrum for 300 realisations.
The MC method matches the experimental conditions closest, since the experiment shows the averaged spectrum of many randomly orientated complexes. To avoid propagating density matrices for many random orientations, the $xyz$ approach can be chosen instead as both spectra show excellent agreement. Neither the integrated dipole moments, nor the fitted ones yield comparable results to the MC method and only in the Markovian limit it is possible to run the propagation with pre-averaged dipole moments. Here, we chose a longer bath correlation time $\gamma^{-1}=166$~fs for the comparison of different rotational averages as the differences vanish for decreasing $\gamma^{-1}$.\\ 
In the following we always use several orientations for rotational averages and do not resort to pre-averaged dipole moments.

\subsection{Broadening of the spectral lines}
\label{sec:broad}

If we compare the only rotationally averaged absorption spectra in figure~\ref{fig:1d-both} (a) with the experimentally measured ones \cite{Cho2005}, we find that the rotationally averaged peaks are too narrow. Furthermore six peaks are well pronounced while in experiments several peaks overlap and the third exciton eigenstate is not individually resolved. This result is expected, since there are several additional effects besides rotational averaging, which broaden the peaks and bring the simulation results closer to the experimental data. If we increase either $\lambda$ (not shown) or the relaxation time $\gamma^{-1}$, the peaks in the spectrum become broader, see figure~\ref{fig:1d-both}~(b). Both parameters change the spectral density and hence the coupling to the vibrational modes. The third mechanism is static disorder caused by the slowly fluctuating protein environment. We model the static disorder by adding a Gaussian distributed noise of a given standard deviation (SD) to each diagonal term in (\ref{eq:H}). The resulting spectra are shown in figure~\ref{fig:1d-averages}~(b) where we obtain broader peaks and show with increasing disorder the diminishing of the third peak in the spectrum. In addition we observe that the broadening changes the relative peak heights of the second and fourth peak yielding a better agreement with the experimental results.

For the absorption spectrum changing the spectral density and inhomogeneous broadening yield similar results and it is not possible to distinguish which mechanism is relevant for the FMO complex. It is in the 2d echo-spectra where the difference becomes visible since the disorder results in a broadening along the diagonal frequencies only, while changing the exciton-phonon coupling broadens the peaks in all directions \cite{Cho2008}.

\section{Two-dimensional echo-spectroscopy}
\label{sec:2d}

\begin{figure}[t]
\begin{center}
\includegraphics[width=0.7\textwidth]{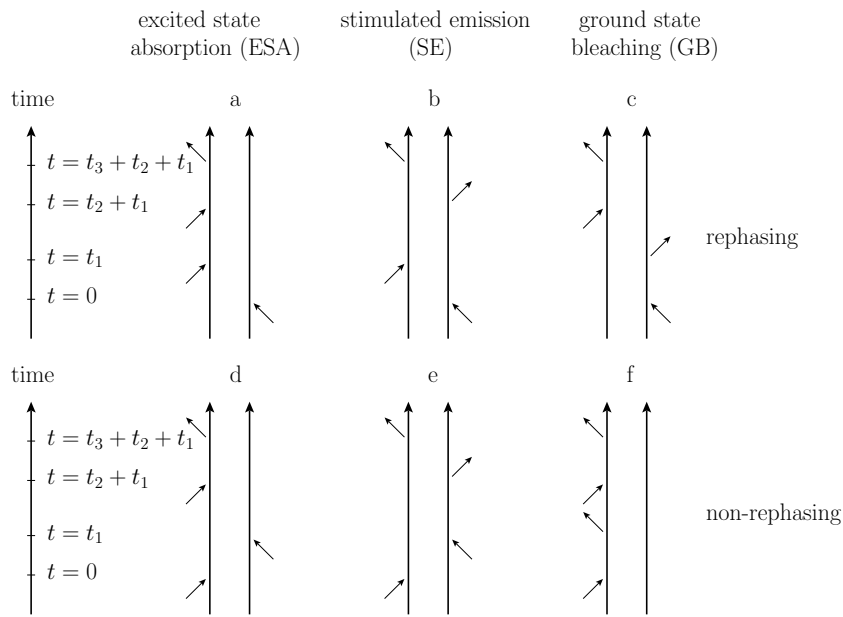}
\end{center}
\caption{The six Feynman diagrams included in the 2d echo-spectrum calculation. Vertical axis represent the time propagation and the action of the dipole operators are shown by diagonal arrows. For a complete theoretical description see \cite{ShaulMukamel}.}
\label{fig:Feynmnan}
\end{figure}

In addition to the information provided by absorption spectroscopy, two-dimensional echo-spectroscopy is used to study time-resolved processes such as energy transfer or vibrational decay as well as to measure intermolecular coupling strengths \cite{Tanimura2009,Cho2006,Abramavicius2008}.  Two-dimensional echo-spectroscopy consists of a sequence of three ultra-short pulses and a resulting signal pulse with fixed time-delay between the second and third pulse \cite{Read2009,Brixner2004}. The theory of 2d echo-spectroscopy is explained in detail in \cite{Cho, ShaulMukamel,Hamm05}. 
The 2d echo-spectra are measured in two configurations which correspond to the rephasing (RP) and non-rephasing (NR) contributions of the third order response function 
\begin{equation}\label{eq:s3}
\fl
S^{(3)}(t_3,t_2,t_1) = \left(\frac{i}{\hbar} \right)^3 \Theta(t_3)\Theta(t_2)\Theta(t_1)\, \tr\left(\mu\left(t_3+t_2+t_1\right) \left[\mu \left(t_2+t_1\right),\left[\mu\left(t_1\right),\left[  \mu (0),\rho_0\right]\right]\right] \right).
\end{equation}
Using the impulsive approximation which assumes $\delta$-functions for the temporal envelopes of the pulses, the two components of the spectrum corresponding to $S^{(3)}(t_3,t_2,t_1)=S_{\rm RP}^{(3)}(t_3,t_2,t_1)+S_{\rm NR}^{(3)}(t_3,t_2,t_1)$ read
\begin{eqnarray}
I_{\rm RP}(\omega_3,t_2,\omega_1) &= \int_{-\infty}^{\infty}   dt_1 \int_{-\infty}^{\infty} dt_3 \rme^{\rmi \omega_3 t_3- i \omega_1 t_1} S_{\rm RP}^{(3)}(t_3,t_2,t_1)  
\label{eq:I_RP}	, 	\\
I_{\rm NR}(\omega_3,t_2,\omega_1) &= \int_{-\infty}^{\infty}   dt_1 \int_{-\infty}^{\infty} dt_3 \rme^{\rmi \omega_3 t_3+ i \omega_1 t_1} S_{\rm NR}^{(3)}(t_3,t_2,t_1)
\label{eq:I_NR} 
.\end{eqnarray}
Sorting this expression for the rephasing and the non-rephasing parts yields six different pathways, which represent the stimulated emission, the ground-state bleach and the excited state absorption \cite{Cho, ShaulMukamel, Hamm05, Chen2010}, shown schematically in figure~\ref{fig:Feynmnan}. 
By taking the time between second and third pulse as a control parameter as done in (\ref{eq:I_RP}) and (\ref{eq:I_NR}) one can study the exciton dynamics which is probed by recording the emitted signal pulse. Due to the excited state absorption the dynamics is more involved compared to the one dimensional case. Including a double excited manifold, we construct an enlarged exciton Hamiltonian $H_{\rme}^{2d}$ which is block diagonal and of the form
\begin{equation}
\label{eq:2d-H}
 H_{\rme}^{2d}=
\left(
\begin{array}{ccc}
 H_{\rme}^{\rm gs} & 0 & 0 \\  
 0 & H_{\rme}^{\rm 1 ex}& 0 \\ 
  0 & 0 & H_{\rme}^{\rm 2 ex} 
\end{array}\right)
,\end{equation}
where $H_{\rme}^{\rm gs}=0$ represents the ground state, $H_{\rme}^{\rm 1 ex}$ is the single-exciton Hamiltonian with $N_{\rm sites}^2$ elements as defined in (\ref{eq:H}), and the energy levels of the two excitons are described by $H_{\rme}^{\rm 2 ex}$. The two-exciton Hamiltonian $H_{\rme}^{2d}$ has ${\left[N_{\rm sites}(N_{\rm sites}-1)/2\right]}^2$ entries which are constructed from the matrix elements of $H_{\rme}^{\rm 1 ex}$ following \cite{Cho, Cho2005}. For the FMO complex $H_{\rme}^{2d}$ is a $29\times29$ matrix. 
In site basis, we denote the two exciton states by  $\left| ij\right\rangle$, where $0<i<j\le N_{\rm sites}$, and the diagonal entries of $H_{\rme}^{\rm 2 ex}$ are given by 
\begin{equation}
\fl
 \left\langle ij \mid H_\rme^{\rm 2 ex}\mid ij\right\rangle=\left\langle i \mid H_\rme^{\rm 1 ex}\mid i\right\rangle+\left\langle j \mid H_\rme^{\rm 1 ex}\mid j\right\rangle
.\label{eq:H2ex-diag}
\end{equation}
The off-diagonal elements are given by
\begin{eqnarray}
\fl
\left\langle ij \mid H_{\rme}^{\rm 2 ex}\mid kl\right\rangle=
& \delta_{ik}\left(1-\delta_{jl}\right)\left\langle j \mid H_{\rme}^{\rm 1 ex}\mid l\right\rangle %\nonumber \\&
+\delta_{il}\left(1-\delta_{jk}\right)\left\langle j \mid H_{\rme}^{\rm 1 ex}\mid k\right\rangle \nonumber \\
&+\delta_{jk}\left(1-\delta_{il}\right)\left\langle i \mid H_{\rme}^{\rm 1 ex}\mid l\right\rangle%\nonumber \\ &
+\delta_{jl}\left(1-\delta_{ik}\right)\left\langle i \mid H_{\rme}^{\rm 1 ex}\mid k\right\rangle
\label{eq:H2ex-offdiag}
.\end{eqnarray}
Physically, this means that the $|ij\rangle$ and $|kl\rangle$ two-exciton states are coupled if they have one exciton in common while a second exciton is shifted from one site to another. The coupling constant is the same as for the transition of the corresponding exciton shift in the one exciton dynamics. Note also that the linear coupling to the independent baths, each of them associated to one BChl, leads to couplings between the one exciton state $\left| i \right\rangle$ and the two exciton states  $\left| i j \right\rangle$ which do both influence the bath at site $i$ and are in return influenced by the bath correlations. 

Following \cite{May2004}, we construct the dipole operator associated with the extended Hamiltonian (\ref{eq:2d-H}) 
\begin{equation}
 \label{eq:2d-mu}
 \mu=\left(\begin{array}{ccc}
 0 & \mu^{-}_{\rm 1ex} & 0 \\  
 \mu^{+}_{\rm 1ex} & 0& \mu^{-}_{\rm 2ex}  \\ 
  0 & \mu^{+}_{\rm 2ex}  & 0 
\end{array}\right)
.\end{equation}
The zero to single excitation operators $\mu^{\pm}_{\rm 1 ex}$ are defined as in (\ref{eq:1d-mu}) and (\ref{eq:mu_1ex}). The single to double excitation operators are defined according to $\left\langle ij\mid\mu^{+}_{\rm 2 ex}\mid k\right\rangle=\delta_{ik} \left\langle j\mid\mu^{+}_{\rm 1 ex}\mid 0\right\rangle+\delta_{jk} \left\langle i\mid\mu^{+}_{\rm 1 ex}\mid 0\right\rangle$ and $\mu^{-}_{\rm 2 ex}=\left({\mu^{+}_{\rm 2 ex}}\right)\transpose$. 

In order to calculate 2d spectra, we have to evolve the $29 \times29$ density matrix to obtain the third order response function $S^{(3)}$. The hierarchical method incorporates the effect of the baths through the auxiliary system of the $\sigma$-matrices. When calculating the 2d spectra, we propagate according to the Feynman diagrams shown in figure~\ref{fig:Feynmnan} and act on the density matrix and the auxiliary matrices with dipole operator to simulate the interaction with the laser field. 

We observe in (\ref{eq:s3}) that the 2d spectra depend on the fourth order dipole moments. To take into account the random orientation of the samples we need to compute rotational averages $\left\langle \mu_{i}^{2}\mu_{j}^{2}\right\rangle_{\rm rot}$ which are performed by sampling $10$ laser polarisation-vectors aligned to the vertices of a dodecahedron in one half of the coordinate space. We take the vertices $\left( \pm 1,\pm 1,1\right)$,  $\left( 0,\pm \frac{1}{\phi},\phi\right)$, $\left( \pm \frac{1}{\phi},\phi,0\right)$ and $\left(\pm\phi, 0, \frac{1}{\phi}\right)$, where $\phi=\frac{1}{2}\left(1+\sqrt{5}\right)$ denotes the golden ratio. We have compared the dodecahedral summation with the full spherical averaging using a $10000$-shoot MC and obtain differences of less than $1$~percent of the MC results.

\subsection{Numerical results for the 2d echo-spectrum of the FMO complex}

For the simulation of 2d spectra, we calculate the third order response function $S^{(3)}(t_1,t_2,t_3)$ in (\ref{eq:s3}). We obtain $128^{2}$ data points with a spacing of $16$~fs in $t_{1}$ and $t_{3}$ direction using a numerical time step of $2$~fs. This means that we have to perform over $131~000$ propagation steps for each of the six Feynman diagrams. Depending on the delay time $T_{\rm delay}=t_{2}$ our simulations take 48 to 53 minutes on a single NVIDIA Fermi C2050 GPU which provides a $38$ fold speed up compared to the CPU algorithm with filtering methods that reduce the hierarchy \cite{Chen2011}. We always show the sum of the real parts of the rephasing and the non-rephasing spectrum.
\begin{figure}[t]
\includegraphics[width=\textwidth]{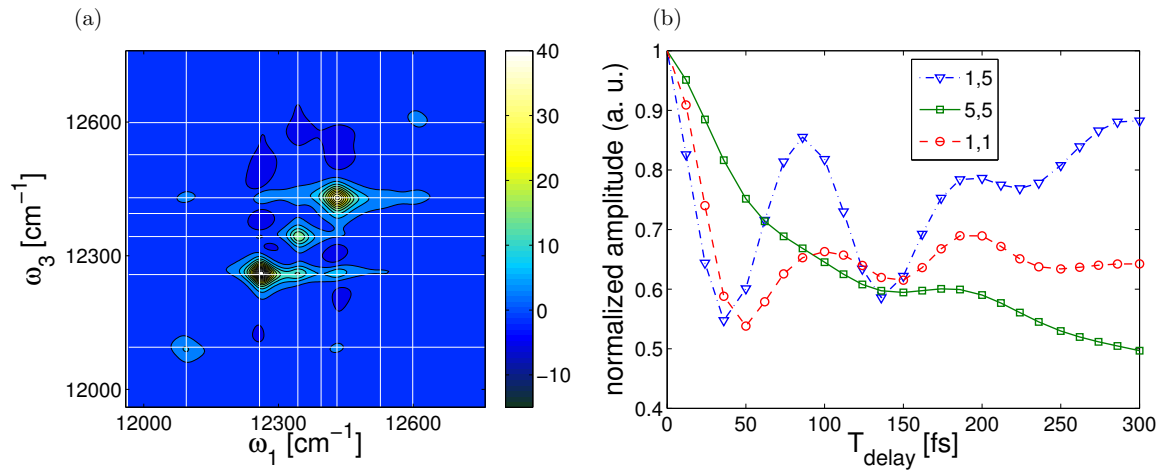}
\caption{(a) Rotationally averaged 2d spectrum $Re[I_{\rm NR}+I_{\rm RP}]$ of the FMO complex calculated using the HEOM with  $\lambda=35$~cm$^{-1}$, $T=77$~K and  $\gamma^{-1}=50$~fs and a delay time $T_{\rm delay}=0$~fs. (b) Oscillations as a function of the delay time for the $(1,1)$ and $(5,5)$ diagonal peak and the $(1,5)$ peak below the diagonal. The points have been connected to guide the eye.}
\label{fig:HEOM2d-DodeAverage}
\end{figure}

Figure \ref{fig:HEOM2d-DodeAverage}~(a) shows a simulated 2d spectrum for a rotationally averaged sample with no disorder at $T=77$~K. 
The figure is encoded using a linear colour scale. We divided the raw data by a constant to obtain an amplitude on a $-10$--$40$ range.
The thin white lines have been added to indicate the expected peak positions. The peaks are denoted by their excitonic energy-level indices $(i,j)$, with the first number referring to the $\omega_3$-axis and the second one to the $\omega_1$-axis. The spectrum shows similarities to the experimentally measured one in \cite{Cho2005} insofar as it reproduces the large peak on the diagonal at the second eigenstate and a second near the fourth and fifth. Furthermore we find negative amplitudes above the diagonal and positive peaks below. As expected from the discussion of the absorption spectrum in section \ref{sec:broad}, the peak shapes differ from the experimental ones due to the lack of static disorder.

Figure~\ref{fig:HEOM2d-DodeAverage}~(b) shows the peak oscillations for the diagonal peaks $(1,1)$ and $(5,5)$ and the off-diagonal peak $(1,5)$ below the diagonal.We renormalised the peak heights at $T_{\rm delay}=0$~fs to unity such that the time resolved amplitude of the three peaks are conveniently put into the same panel. We choose these peaks because figure~\ref{fig:HEOM2d-DodeAverage} shows peak $(1,5)$ as the highest off-diagonal peak in the line where $\omega_{3}$ equals the lowest lying eigenstate. We see that diagonal and off-diagonal peaks oscillate in time. This indicates that our model imitates the coherent energy transport within the FMO complex that has been observed experimentally \cite{Engel2007a,Panitchayangkoon2010}. Theoretically, an off-diagonal peak $(i,j)$ is expected to undergo oscillations with a frequency corresponding to the difference of its excitonic eigenenergies $(E_j-E_i)$. For the $(1,5)$ peak we expect a period of $99$~fs which is in good agreement to our numerical result yielding a best-fit period of $97$~fs. When we calculate the peak heights, we integrate over squares of width $\Delta \omega=16$~cm$^{-1}$ centred around the white crosses. We follow the changes of the peak heights with increasing delay time by plotting the absolute value of the spectrum normalised with respect to the value at $T_{\rm delay}=0$~fs. 
\begin{figure}[t]
\includegraphics[width=\textwidth]{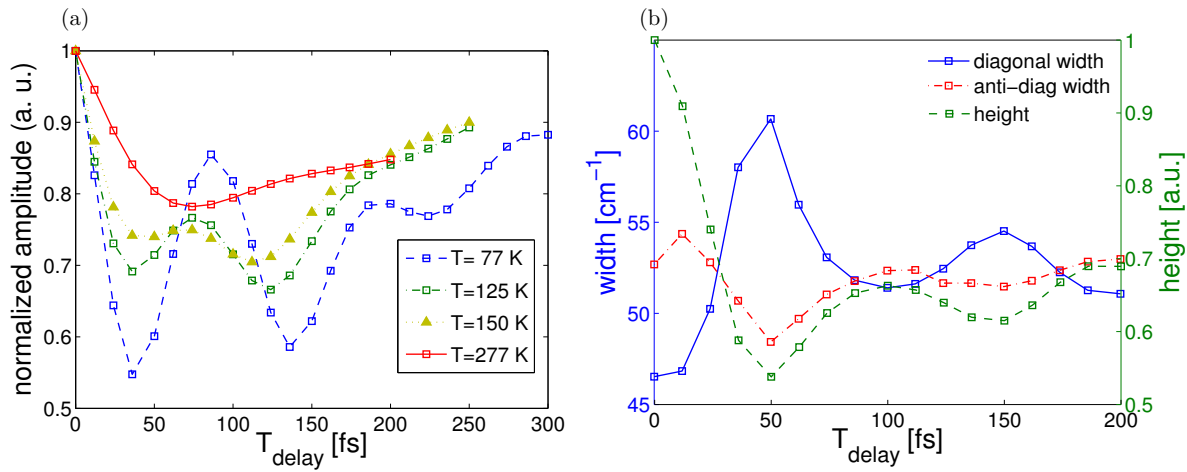}
\caption{(a) Amplitude of the cross peak $(1,5)$ of the rotationally averaged spectra of the FMO complex as a function of delay time for different temperatures. Temperature dependence of the peak oscillations for the $(1,5)$ cross peak. (b) Oscillations in the peak width of the $(1,1)$ diagonal peak measured in diagonal and anti-diagonal direction compared to the peak height at $T=77$~K. The points have been connected to guide the eye.}
\label{fig:HEOM2d-noDisorder-peak-oscillations}
\end{figure}

In order to analyse the robustness of the peak oscillations when the thermal fluctuations increase we plot in figure \ref{fig:HEOM2d-noDisorder-peak-oscillations}~(a) the oscillations of the $(1,5)$ cross peak for several temperatures. We compare with the experimental data in \cite{Panitchayangkoon2010} and observe a qualitative similar behaviour with weaker oscillations and faster decay when temperature is increased. We do not obtain the picoseconds lasting beatings that have been observed experimentally. One possible factor is the shape of the spectral density that is used for the HEOM. We fit the oscillations in figure~\ref{fig:HEOM2d-noDisorder-peak-oscillations}~(a) by an exponentially decaying sine.
The fit is done separately for the stimulated emission and excited state absorption pathway, which couple to one or two baths respectively. Based on the damping of the oscillations on peak $(1,5)$ the temperature dependence of the dephasing rate of our simulation is given by ${\gamma_{\rm deph}^{\rm ESA}\left(T\right)}/{T}=(0.66 \pm 0.05)$~cm${}^{-1}$/K and ${\gamma_{\rm deph}^{\rm SE}\left(T\right)}/{T}=(0.50 \pm 0.03)$~cm${}^{-1}$/K.
This is in good agreement to what we expect for the chosen spectral density ($2 k_B \lambda/(\gamma T)=0.46$~cm$^{-1}$/K) and the value $0.52$~cm$^{-1}$/K determined from the experimentally measured data \cite{Panitchayangkoon2010}.  

In the experiments an anti-correlation in peak shapes between the diagonal peak width and the height \cite{Engel2007a}, or alternatively the ratio of the diagonal peak-width over the anti-diagonal height \cite{Collini2010a} has been found. The width was measured by fitting a Gaussian to a diagonal cut through the spectrum and taking its standard deviation. Applying the same procedure to the simulation data yields also anti-correlations shown in figure~\ref{fig:HEOM2d-noDisorder-peak-oscillations}~(b). 

 \begin{figure}[t]
\includegraphics[width=\textwidth]{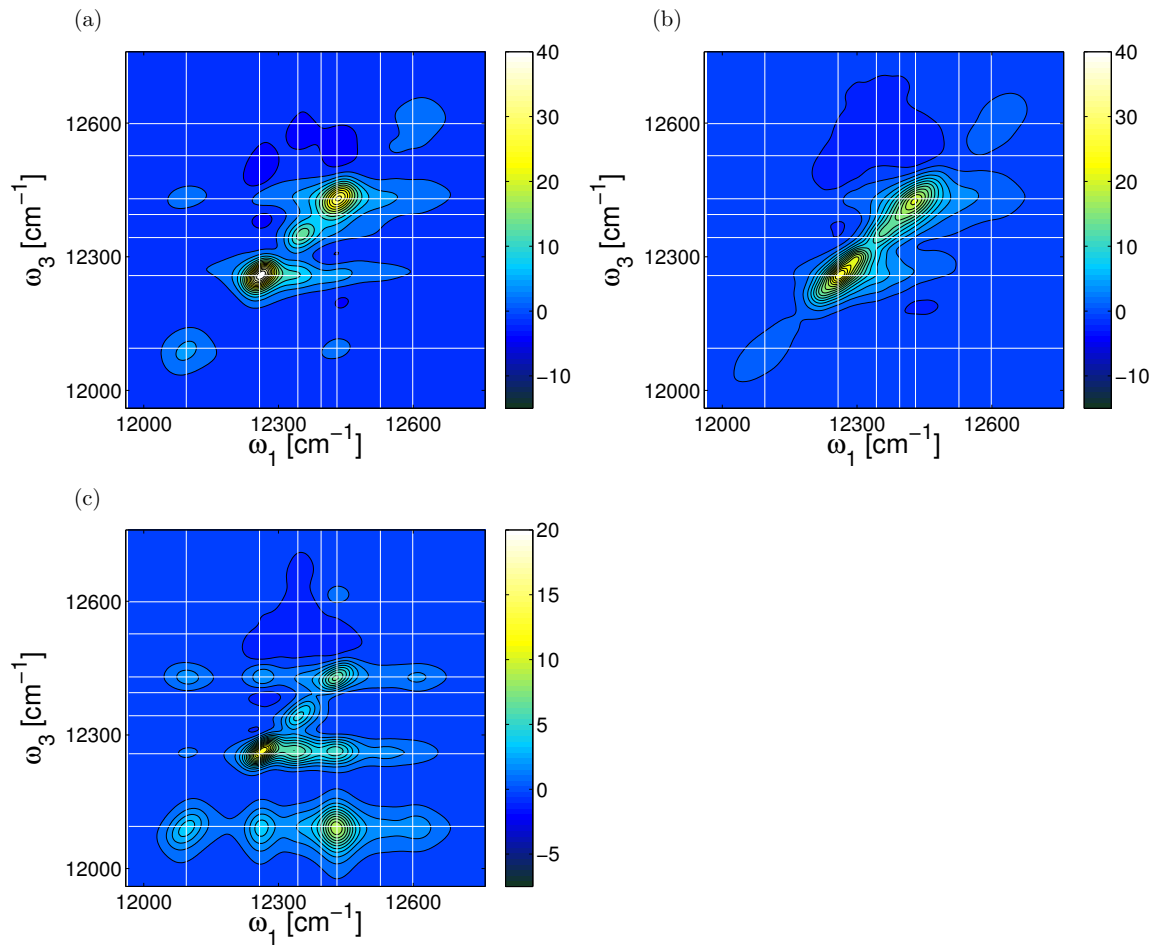}
\caption{Two-dimensional echo-spectra $Re[I_{\rm NR}+I_{\rm RP}]$ of the FMO complex with rotational and disorder averages at temperature $T=77$~K. The averages are performed over $400$ realisations. The spectral-density parameters are $\lambda=35$~cm$^{-1}$ and $\gamma^{-1}=50$~fs. (a) Gaussian disorder with standard deviation (SD) $25$~cm$^{-1}$ at delay time $T_{\rm delay}=0$~fs. (b) SD$=50$~cm$^{-1}$, $T_{\rm delay}=0$~fs. (c) SD$=25$~cm$^{-1}$, $T_{\rm delay}=4000$~fs. 
}
\label{fig:HEOM2d-with-average}
\end{figure}

Next, we analyse the role of static disorder in the results for the 2d echo-spectra. When including static disorder, the peaks get elongated along the diagonal as observed when comparing the spectra at zero delay time shown in figures~\ref{fig:HEOM2d-DodeAverage}~(a) and \ref{fig:HEOM2d-with-average}~(a) and (b) for increasing values of disorder. The plots with disorder show better agreement with the experiment \cite{Cho2005}. Figure~\ref{fig:HEOM2d-with-average}~(c) shows the 2d spectra for SD$=25$~cm$^{-1}$ for a long delay time $T_{\rm delay}=4000$~fs for which the system approaches its thermal stationary state. We observe that the excitation moves from the diagonal peaks into the regions with lower frequency $\omega_{3}$. This occurs as the system relaxes to lower lying eigenstates during the delay time. In experiments by Brixner \etal \cite{Brixner2005} the decay of the population into the ground state of the system is also visible.

The absence of the experimentally seen oscillation in the HEOM calculation in Ref.~\cite{Chen2011} has led Chen \etal to conclude that the importance of static disorder requires careful reconsideration. First-principles simulations of static disorder in the FMO complex are computationally extremely time-consuming, since several hundred of realisations are required to reach converged results. In order to study the role of static disorder in the oscillations of the peaks of the spectrum and to ensure numerical convergence, we construct a dimer to model the $(1,5)$ peak of the FMO spectrum. For the model dimer we chose the Hamiltonian such that the first and fifth FMO complex eigenvalues are reproduced. The coupling between both sites is set to $50$~cm$^{-1}$.
\begin{figure}[t]
\includegraphics[width=\textwidth]{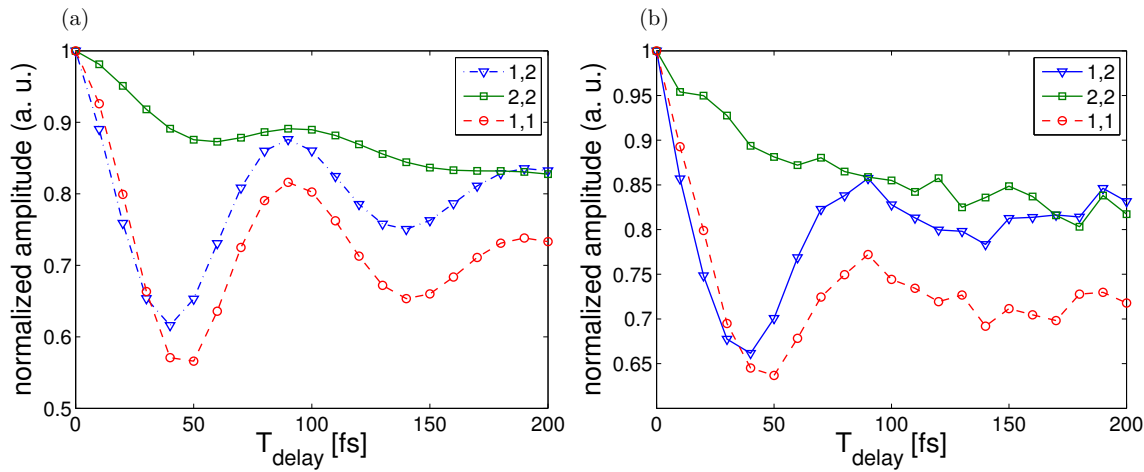}
\caption{Peak oscillations in a model dimer for $T=77$~K, $\lambda=35$~cm$^{-1}$ and $\gamma^{-1}=50$~fs. To concentrate on the disorder effects, no rotational average is performed and the calculation is done using $\mu_{1}=-5.4$ and $\mu_{2}=12.2$ (a) no static disorder, (b) Gaussian disorder with standard deviation (SD) $25$~cm$^{-1}$. The average was done over $1200$ realisations at each delay time. The points have been connected to guide the eye.}
\label{fig:Dimer-osc}
\end{figure}
The simulations with and without disorder show oscillations on the diagonal and off-diagonal peaks. The most striking effect of disorder is a decreased oscillatory amplitude. While the cross peak in the simulation without disorder has lost $45$~percent of its amplitude in the first minimum, in the disorder case the loss is only $35$~percent. From a fit, we observe that the periods of the disorder curves are slightly shorter $T=92$~fs and $T=98$~fs for the $(1,1)$ and $(2,1)$ curves while in the no-disorder plot $T=95$~fs and $T=98$~fs has been obtained, respectively. 
For the FMO complex we expect that oscillations are reduced in amplitude by a similar magnitude and therefore still persist within the model.

\section{Conclusions}
\label{sec:conclusions}

We have presented calculations of the absorption spectra and the 2d echo-spectra for the FMO complex at different temperatures using the HEOM for solving the exciton and the vibrational bath dynamics and adding static disorder to account for the fluctuating environment. Important ingredients necessary to find good agreement with experiments for the absorption spectra using the HEOM propagation method is to perform the rotational average and to include static disorder. We have also discussed how other simplified propagation methods affect the line shapes and shift the peaks in the spectra.

Our rotationally averaged 2d echo-spectra reflect the main features of the experimentally measured ones. At delay time $T_{\rm delay}=0$ we find negative contributions above the diagonal and elongated peaks in the diagonal which are transferred to the lowest exciton states after long times on the picosecond scale. The HEOM method yields an oscillatory behaviour of peak amplitudes over a wide temperature range. At $T=77$~K we obtain oscillations of cross peak amplitudes up to $300$~fs with the expected frequency that corresponds to the difference in excitonic eigenenergies. In addition oscillations on the diagonal peaks are present. The long duration of oscillations in the experimental data (up to $1.8$~ps in Ref.~\cite{Panitchayangkoon2010}) suggests that further refinements of the theoretical models are necessary, for example by adjusting the form of the spectral density which affects the population dynamics \cite{Nalbach2011}.

Remarkably, we find excellent agreement with the measured damping rate of the oscillations due to thermal fluctuations. This implies a good choice of the parameters of the model and might explain why we obtain different results compared to previous calculations, which were done for a higher value of $\gamma^{-1}=100$~fs \cite{Chen2011}, leading to an increased dephasing by a factor of two. At ambient temperatures the HEOM have the advantage to require less terms to converge while approximate methods such as the secular and full Redfield approximations do not yield the correct time-evolution of the density matrix and overestimate the thermalisation rate.

In agreement with the experimental measurements we observe the anti-correlated oscillations of the peak heights and peak shapes for the diagonal peaks. 
The analysis of a model dimer shows that the cross peak oscillations persist even in the presence of static disorder which mainly results in a reduction of oscillation amplitude by about $30$ percent. 

\ack

This work is supported by the Emmy-Noether programme of the DFG (KR~2889/2), the DAAD project 50240755, the Spanish MINCINN AI  DE2009-0088 (Acciones Integradas Hispano-Alemanas), the Spanish MICINN project FIS2010-18799, and the Ram{\'o}n y Cajal programme. We thank Akihito Ishizaki for helpful discussions. Time on the Harvard School of Engineering and Applied Sciences ``Resonance'' GPU cluster and support by the SEAS Academic Computing team are gratefully acknowledged.

\end{document}